\documentclass{PoS}
\usepackage{amsmath}
\usepackage{subfigure}
\usepackage{xspace}
\newcommand{\Dslash}{\ensuremath{D\hspace{-1.5ex} /}}
\newcommand{\Tr}{\ensuremath{\operatorname{Tr}}}

\newcommand{\einh}[1]{\ensuremath{\,\text{#1}}}
\newcommand{\MeV}{\einh{MeV}}

\def\Eq#1{Eq.~(\ref{#1})}
\def\Fig#1{Fig.~\ref{#1}}
\newcommand{\Phibar}{\ensuremath{\bar{\Phi}}}
\newcommand{\LPQM}{\ensuremath{\mathcal{L}_{\textrm{PQM}}}\xspace}

\newcommand{\muT}{\left(\frac{\mu}{T}\right)}
\newcommand{\onefig}{0.44\linewidth}
\newcommand{\twofigs}{0.44\linewidth}

\newcommand{\pade}{Pad\'e}

\graphicspath{
{./}
{../../figures/}
{./figures/}
}

\title{
 \vspace*{-2.cm}
 \begin{flushright}\texttt{\footnotesize
  BI-TP 2011/33\\}
 \end{flushright}
 \vfill
 Towards finite density QCD with Taylor expansions}

\ShortTitle{Towards finite density QCD with Taylor expansions}

\author{F.~Karsch\\
Physics Department, Brookhaven National Laboratory, Upton, NY 11973, USA\\
Fakult\"at f\"ur Physik, Universit\"at Bielefeld, D-33615
  Bielefeld, Germany\\
 E-mail: \email{karsch@physik.uni-bielefeld.de}, \email{karsch@bnl.gov}}
\author{B.-J.~Schaefer\\
Institut f\"{u}r Physik, Karl-Franzens-Universit\"{a}t Graz,
  A-8010 Graz, Austria\\
Institut f\"{u}r Theoretische Physik, Justus-Liebig-Universit\"{a}t Gie{\ss}en, D-35392 Gie{\ss}en, Germany\\
 E-mail: \email{bernd-jochen.schaefer@uni-graz.at}}
\author{\speaker{M.~Wagner}%
        \\
       Fakult\"at f\"ur Physik, Universit\"at Bielefeld, D-33615
  Bielefeld, Germany\\
       E-mail: \email{mwagner@physik.uni-bielefeld.de}}
 \author{J.~Wambach\\
 Institut f\"{u}r Kernphysik, TU Darmstadt, D-64289
  Darmstadt, Germany\\
  Gesellschaft f\"{u}r
  Schwerionenforschung GSI, D-64291 Darmstadt, Germany\\
  E-Mail: \email{jochen.wambach@physik.tu-darmstadt.de}}

\abstract{We analyze general convergence properties of the Taylor
  expansion of observables to finite chemical potential in the
  framework of an effective 2+1 flavor Polyakov-quark-meson model. To
  compute the required higher order coefficients a novel technique
  based on algorithmic differentiation has been developed. Results for
  thermodynamic observables as well as the phase structure obtained through the
  series expansion up to 24th order are compared to the full model
  solution at finite chemical potential. The available higher order
  coefficients also allow for resummations, e.g. \pade{} series, which
  improve the convergence behavior. In view of our results we discuss
  the prospects for locating the QCD phase boundary and a possible
  critical endpoint with the Taylor expansion method.}

\FullConference{XXIX International Symposium on Lattice Field Theory \\
		 July 10-16, 2011\\
		 Squaw Valley, Lake Tahoe, California}

\begin{document}

%%%%%%%%%%%%%%%%%%%%%%%%%%%%%%%%%%%%%%%%%%%%%%%%%%%%%%%%%%%%%%%%%%%%%%%%%%%%%
%%
%% Introduction
%%
%%%%%%%%%%%%%%%%%%%%%%%%%%%%%%%%%%%%%%%%%%%%%%%%%%%%%%%%%%%%%%%%%%%%%%%%%%%%%
\section{Introduction}

The phase structure of strongly-interacting matter at non-vanishing
temperature and densities and in particular the possible existence of
a critical endpoint (CEP) is currently a very active frontier both
theoretically and experimentally. Our present knowledge of the QCD
phase diagram in the non-perturbative regime rests upon effective
descriptions since first-principle approaches such as QCD lattice
simulations mostly fail at large densities, for recent developments
see e.g.~\cite{levkovalattice2011}. Several extrapolation methods have
been proposed to overcome the fermion sign problem.  All of these
approximations have their own problems and their reliability is still
under investigations, see
e.g.~\cite{Philipsen:2005mjSchmidt:2006usdeForcrand:2010y} for
reviews.  Therefore, it would stand to reason to combine the
approximation schemes with QCD-like effective model calculations. A
comparison of the analytical model results with the lattice
calculations will shed light on their principal applicabilities. Here
we focus on the Taylor expansion
method~\cite{Allton:2002ziAllton:2003vxAllton2005gk}
for small chemical potentials, see also~\cite{Karsch:2010hm}.

%%%%%%%%%%%%%%%%%%%%%%%%%%%%%%%%%%%%%%%%%%%%%%%%%%%%%%%%%%%%%%%%%%%%%%%%%%%%%
%% Taylor expansion
%%%%%%%%%%%%%%%%%%%%%%%%%%%%%%%%%%%%%%%%%%%%%%%%%%%%%%%%%%%%%%%%%%%%%%%%%%%%%

It is based on an expansion of the pressure $p$ in powers of
$x \equiv \mu/T$ around vanishing quark chemical potential~$\mu$:
\begin{equation}
  \label{eq:pressure}
  \frac{p(\mu/T)}{T^4} = \sum_{n=0}^\infty c_n(T) \muT^n\quad \text{with} \quad
  c_n(T) = \left.\frac{1}{n!} \frac{\partial^n\left(p(T,\mu) /T^4
      \right)}{\partial \left(\mu/T\right)^n} \right|_{\mu=0}\ . 
\end{equation}
The relevant Taylor coefficients $c_n (T)$ can be
calculated with standard techniques at $\mu =0$ \cite{Bernard:1996csKarsch:2000psAliKhan:2001ek}.
The convergence of the series may be  improved by a
resummation based on a Pad\'e approximation
\begin{equation}
  [L/M]\equiv
  R_{L,M}(x)=\frac{p(x)}{q(x)}=\frac{p_0+p_1x+\cdots+p_Lx^L}{1+q_1
    x+\cdots+q_Mx^M}\ ,
\end{equation}
where the \pade{} coefficients $p_i$ and $q_j$ can be extracted from
the Taylor coefficients $c_n$ up to the order $L+M$. Hence, no further
input is required for the application of a \pade{}
resummation.  By means of the \pade{} resummation often an extended
convergence range and more stable results with fewer coefficients can
be obtained, in particular in the presence of singularities.

%%%%%%%%%%%%%%%%%%%%%%%%%%%%%%%%%%%%%%%%%%%%%%%%%%%%%%%%%%%%%%%%%%%%%%%%%%%%%
%% PQM
%%%%%%%%%%%%%%%%%%%%%%%%%%%%%%%%%%%%%%%%%%%%%%%%%%%%%%%%%%%%%%%%%%%%%%%%%%%%%
\section{A model analysis}

As an effective QCD-like model framework we chose the $N_f=2+1$ flavor
Polyakov quark-meson (PQM) model which exhibits a chiral crossover at
vanishing densities at $T_\chi \sim 206\MeV$ and a CEP around
$(T_c,\mu_c) \sim (185,167)\MeV$.
Its thermodynamics at $\mu=0$ is in agreement with recent
lattice simulations~\cite{Schaefer:2009ui}.  The PQM model Lagrangian consist of a
quark-meson part 
\begin{equation}
  \label{eq:lpqm}
  \LPQM = \bar{q}\left(i \Dslash - h \phi_5 \right) q + \mathcal{L}_m -\mathcal{U} (\Phi, \Phibar)\ ,
\end{equation}
where the $N_f$ quark fields $q$ are coupled via a Yukawa interaction $h$ to
the scalar and pseudoscalar meson nonets $\phi$ and to the Polyakov loop via the covariant
derivative. The purely mesonic contribution reads
\begin{eqnarray}
\label{eq:mesonL}
  \mathcal{L}_m &=& \Tr \left( \partial_\mu \phi^\dagger \partial^\mu
    \phi \right)
  - m^2 \Tr ( \phi^\dagger \phi) -\lambda_1 \left[\Tr (\phi^\dagger
    \phi)\right]^2  - \lambda_2 \Tr\left(\phi^\dagger \phi\right)^2\nonumber \\
  &&
  +c   \left(\det (\phi) + \det (\phi^\dagger) \right) +
  \Tr\left[H(\phi+\phi^\dagger)\right]\ . 
\end{eqnarray}
The pure gauge sector is encoded (in the logarithmic version of)
the Polyakov loop potential~\cite{Roessner:2006xn}
\begin{eqnarray} 
\label{eq:ulog}
\frac{\mathcal{U}_{\text{log}}}{T^{4}} &=& -\frac{a(T)}{2} \Phibar \Phi+ b(T) \ln \left[1-6 \Phibar\Phi + 4\left(\Phi^{3}+\Phibar^{3}\right)
  - 3 \left(\Phibar \Phi\right)^{2}\right]\ 
\end{eqnarray}
whose parameters are fitted to lattice data. The
matter back-reaction to the pure Yang-Mills system~\cite{Schaefer:2007pw} is not considered in this work. The remaining model
parameters are chosen to reproduce meson masses and decay constants in
the vacuum~\cite{Schaefer:2008hk}.  The grand potential in mean-field
approximation is obtained by integrating the quark loop which yields a
function of the (non-strange and strange) chiral order parameters
($\sigma_x$,$\sigma_y$) and the Polyakov-loop variables
($\Phi,\Phibar$)
\begin{equation*}
  \label{eq:grandpot}
  \Omega = U \left(\sigma_{x},\sigma_{y}\right) +
  \Omega_{\bar{q}{q}} \left(\sigma_{x},\sigma_{y},
    \Phi,\Phibar \right) + 
  \mathcal{U}\left(\Phi,\Phibar\right) \ .
\end{equation*}
Their temperature and chemical potential dependence is found by
minimizing the grand potential
\begin{equation}
  \label{eq:pqmeom}
  \left.\frac{ \partial \Omega}{\partial 
      \sigma_x} = \frac{ \partial \Omega}{\partial \sigma_y}  = \frac{
      \partial \Omega}{\partial \Phi}  =\frac{ \partial
      \Omega}{\partial \Phibar} 
  \right|_{min} = 0\ ,
\end{equation}
where {\em min} labels the global minimum of the potential. 
This introduces an only numerically invertible implicit
$\mu$-dependence in the pressure $p(T,\mu) = -\left
  . \Omega(T,\mu)\right |_{min}$.

%%%%%%%%%%%%%%%%%%%%%%%%%%%%%%%%%%%%%%%%%%%%%%%%%%%%%%%%%%%%%%%%%%%%%%%%%%%%%
%% AD
%%%%%%%%%%%%%%%%%%%%%%%%%%%%%%%%%%%%%%%%%%%%%%%%%%%%%%%%%%%%%%%%%%%%%%%%%%%%%
%%%%%%%%%%%%%%%%%%%%%%%%%%%%%%%%%%%%%%%%%%%%%%%%%%%%%%%%%%%%%%%%%%
%% Runtime
\begin{figure}
\begin{center}
\includegraphics[width=\onefig]{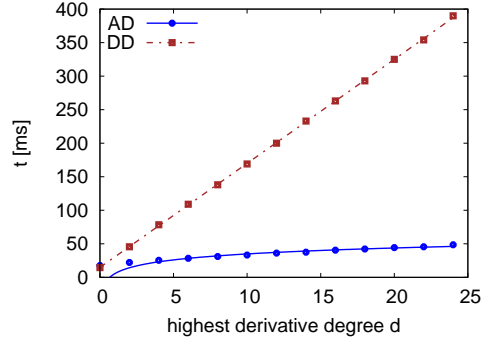}
\caption{Runtime comparison for the DD and AD methods}
\label{fig:mwad_runtime}
\end{center}
\end{figure}

An analytic evaluation of the Taylor coefficients is not only
cumbersome due to the exponentially increasing number of terms but
becomes finally impossible by the implicit $\mu$-dependence. Standard
numerical derivate techniques such as divided differences (DD) fail
at higher orders due to the increasing numerical errors.
By means of the novel derivative technique, based on \emph{algorithmic
  differentiation} (AD), higher order derivatives can be calculated
with extremely high precision, essentially limited only by machine
precision. This works even in the case of only numerically solvable
implicit dependences, such as in \Eq{eq:pqmeom}. Furthermore, also the
performance of the AD technique is superior to the DD method as
demonstrated in \Fig{fig:mwad_runtime} where the time to calculate the
$d$-th order derivatives is shown as a function of $d$. With the DD
method one needs at least $d+1$ function evaluations for a $d$-th
order derivatives and a linear $d$-dependence is obtained in contrast
to the logarithmic dependence for the AD method, see
\cite{Wagner:2009pm} for further details.  High-order Taylor
coefficients calculated with the AD technique in the PQM model are
given in~\cite{Schaefer:2009stWambach:2009ee}. The method is neither
restricted to vanishing $\mu$ nor mean-field approximation and can be
used to search for signals of the CEP in higher
moments~\cite{SchaeferWagner}.

%%%%%%%%%%%%%%%%%%%%%%%%%%%%%%%%%%%%%%%%%%%%%%%%%%%%%%%%%%%%%%%%%%%%%%%%%%%%%
%%
%% Thermodynamics
%%
%%%%%%%%%%%%%%%%%%%%%%%%%%%%%%%%%%%%%%%%%%%%%%%%%%%%%%%%%%%%%%%%%%%%%%%%%%%%%
\subsection{Thermodynamics}

%%%%%%%%%%%%%%%%%%%%%%%%%%%%%%%%%%%%%%%%%%%%%%%%%%%%%%%%%%%%%%%%%%
%% Susceps
\begin{figure}
\centering
\subfigure[$\mu/T = 0.8$]{\includegraphics[width=\twofigs]{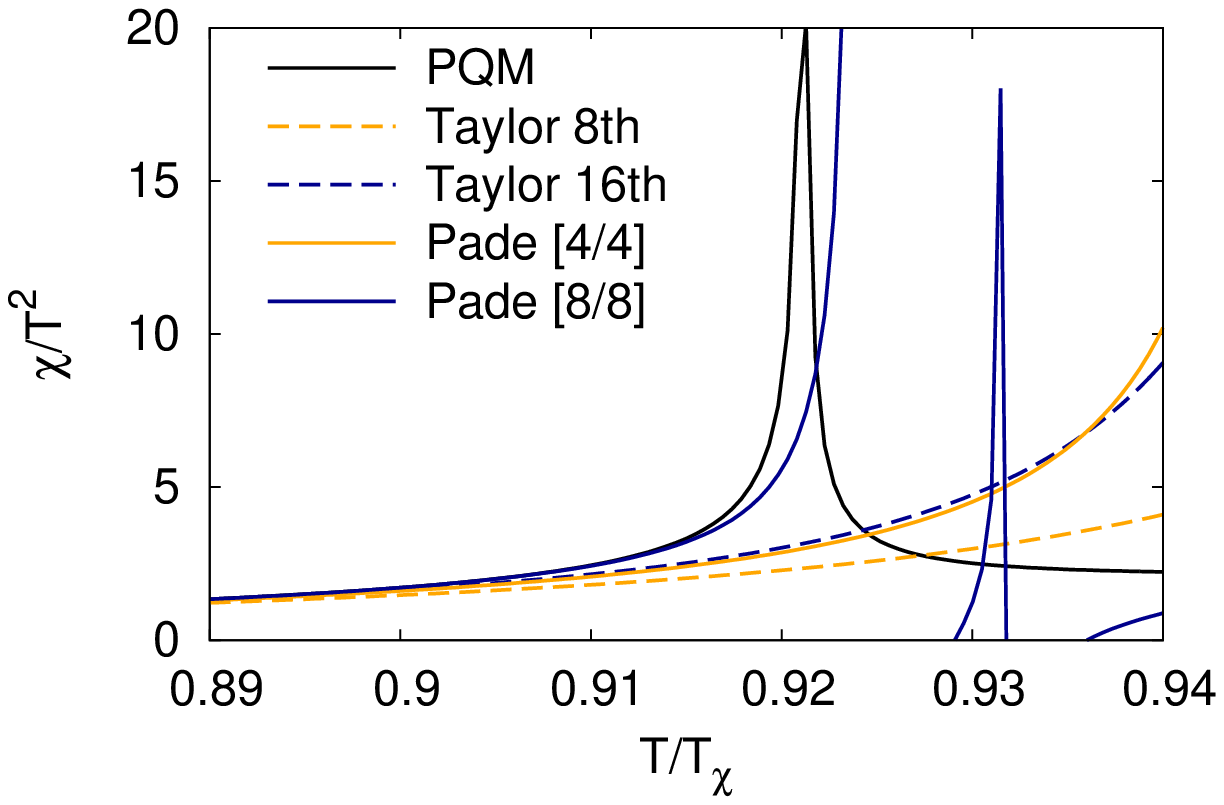}}
\hfill
\subfigure[$\mu/T = 0.9$]{\includegraphics[width=\twofigs]{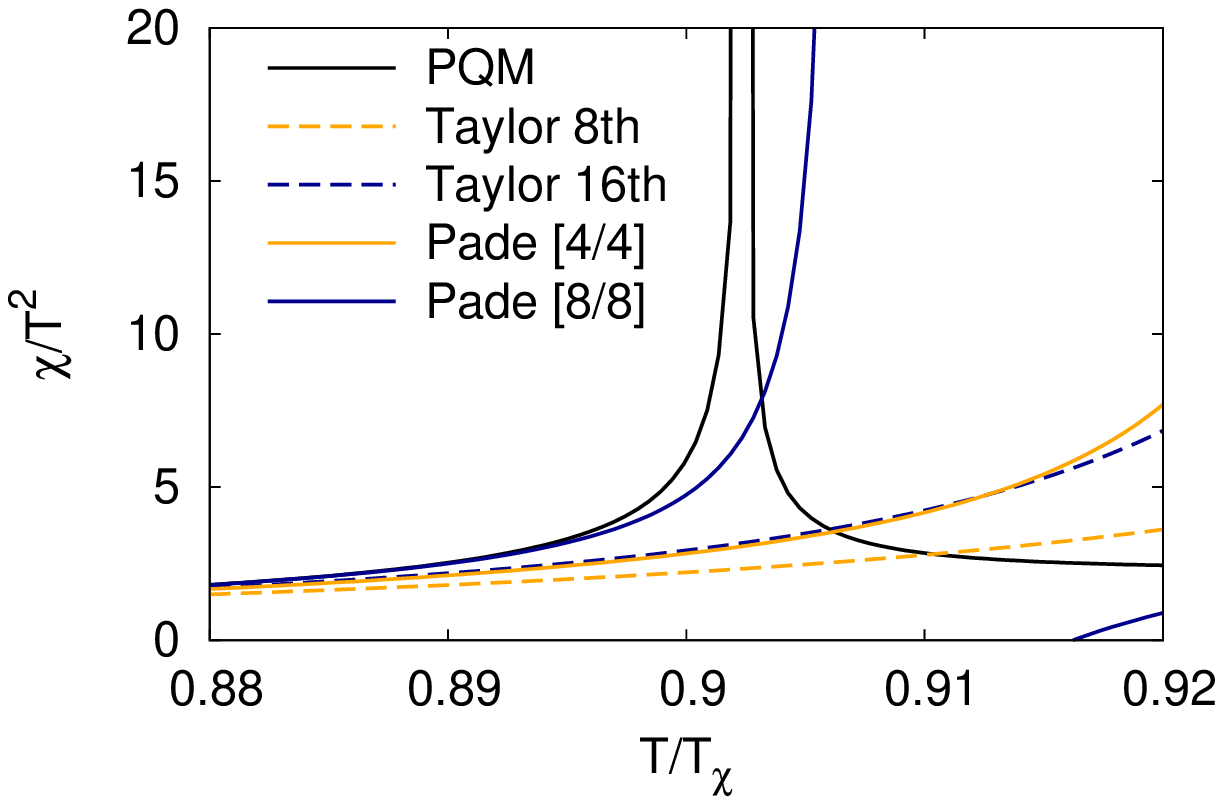}}
\caption{\label{fig:suscep}Comparison of the quark number
  susceptibility in the mean-field model (PQM) with different orders
  of the extrapolation methods in the crossover region (left panel)
  and at the CEP (right panel).}
\label{fig:taylorsuscep}
\end{figure}

The Taylor expansion can be applied straightforwardly to extrapolate
thermodynamic quantities such as the quark-number susceptibility $\chi
= \partial^2 p/(\partial \mu^2)$ to finite densities.  In
\Fig{fig:taylorsuscep} the scaled susceptibility $\chi/T^2$ calculated
in the PQM model is compared to results obtained with the Taylor
expansion and the \pade{} resummation. It diverges exactly at the CEP
which is located for our chosen parameters at $(T_c,\mu_c) \sim (185,
167)\MeV$, correspondingly $\mu_c/T_c \sim 0.9$ (right panel) and is
finite otherwise. Both, the Taylor expansion and the \pade{}
resummation cannot capture this divergence and yield finite values
around $T_c$. For comparison the left panel shows $\chi$ in the
crossover regime ($\mu/T=0.8$).  In the chirally broken phase, i.e.,
for temperatures smaller than the one of the $\chi$ peak, the
agreement improves with increasing expansion orders. At larger
temperatures $T \sim T_\chi$ the extrapolations exhibit some
unphysical peaks which are not present in the model calculation. These
peaks, which are also seen in extrapolations where lattice simulation
coefficients have been used (see e.g.~\cite{Schmidt:2009qq}), are
solely an artifact of the extrapolation. However, an important
observation is that in both cases the $[4/4]$ \pade{} approximation
yields nearly the same result as the $16$-th order Taylor expansion
although it requires only the $8$-th order Taylor coefficients. Hence,
in our case the \pade{} approximation reduces the number of required
derivatives roughly by a factor of two.

%%%%%%%%%%%%%%%%%%%%%%%%%%%%%%%%%%%%%%%%%%%%%%%%%%%%%%%%%%%%%%%%%%%%%%%%%%%%%
%%
%% Phase Boundary
%%
%%%%%%%%%%%%%%%%%%%%%%%%%%%%%%%%%%%%%%%%%%%%%%%%%%%%%%%%%%%%%%%%%%%%%%%%%%%%%
\subsection{Determining the phase boundary and locating the CEP}

The Taylor expansion is limited by the closest singularity in the
complex $\mu$-plane. Its distance to the expansion point can be
obtained from the convergence radius of the series:
\begin{equation}
  r = \lim_{n \rightarrow \infty} r_{2n} = \lim_{n \rightarrow \infty}
  \left|\frac{c_{2n}}{c_{2n+2}} \right|^{1/2}\ . 
\label{eq:convr}
\end{equation}
If the limiting singularity lies on the real axis the convergence
radius corresponds to the location of a critical point.
As the limit $n
\rightarrow \infty$ cannot be carried out in a numerical study the
estimators of the convergence radius at finite $n$ may depend on the
considered observable and approach the true convergence radius
differently. Therefore we also consider the convergence radius for the
quark number susceptibility $r_{2n}^\chi$ which is related to the one of the pressure via
\begin{equation}
  \label{eq:convradchi}
  r_{2n}^\chi = \left|\frac{c^\chi_{2n}}{c^\chi_{2n+2}} \right|^{1/2}=
  \left(\frac{(2n+2)(2n+1)}{(2n+3)(2n+4)}\right)^{1/2}r_{2n+2} \ .
\end{equation}

\begin{figure}[tb]
\centering
\subfigure[\label{fig:convrad}]{\includegraphics[width=\twofigs]{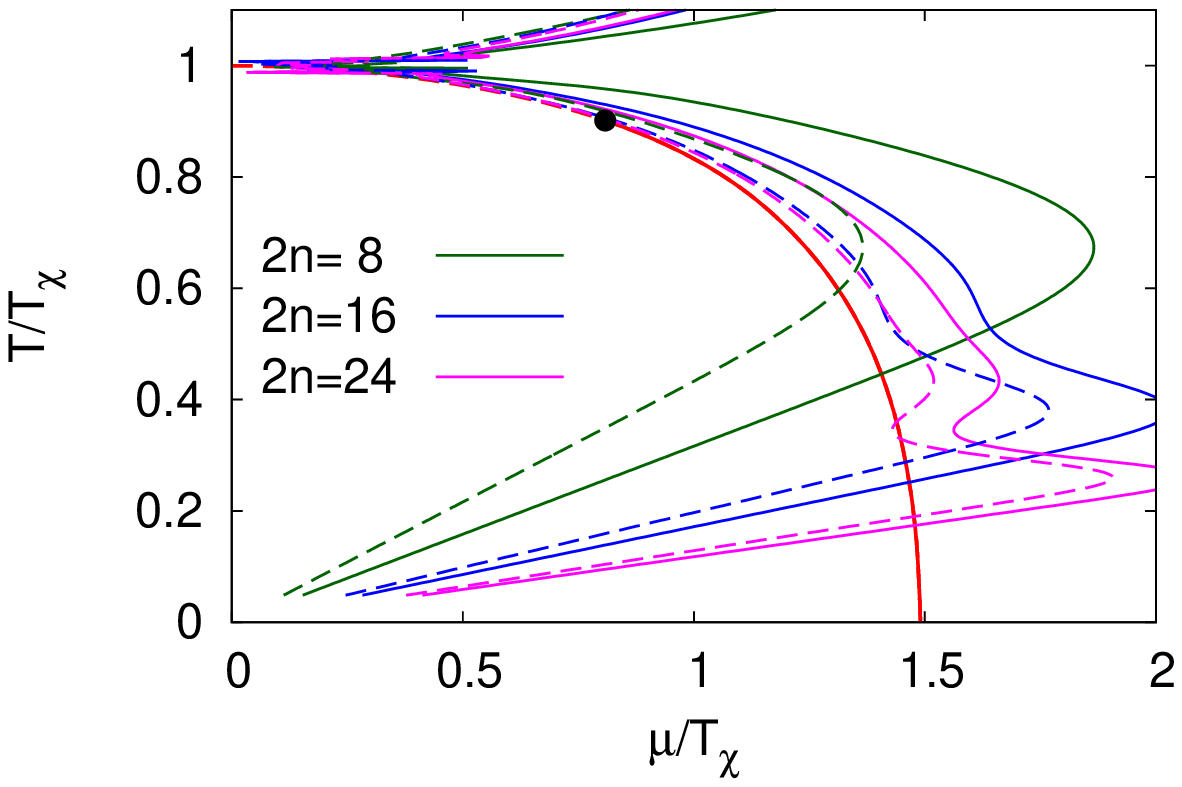}}
\hfill
\subfigure[\label{fig:pbpade}]{\includegraphics[width=\twofigs]{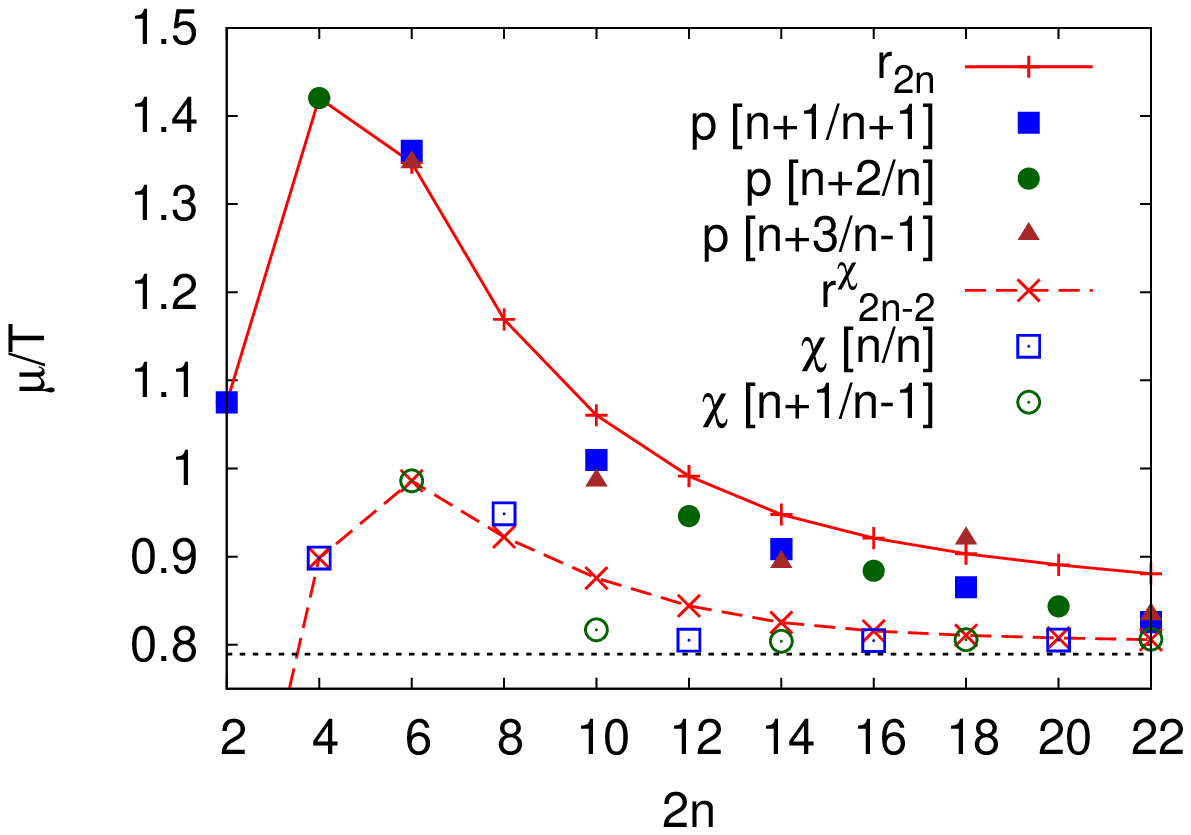}}
\caption{\textbf{Left:} Estimates for the radius of convergence obtained from
  $r_{2n}$ (solid lines) and $r_{2n-2}^\chi$ (dashed lines) for
  different orders of the Taylor expansion. Also shown
  is the chiral phase boundary (red line; dashed:
  crossover, solid: first order). The black dot indicates the CEP.
\textbf{Right:} Estimate of the phase boundary at
  $T=190\MeV \sim0.92\ T_\chi$ with $r_{2n}$ and $r^\chi_{2n}$ and
  poles in the \pade{} approximation of the pressure and quark number
  susceptibility as a function of the order of the Taylor expansion
  $2n$. The used highest coefficient is $c_{2n+2}$. The dotted
  horizontal line indicates the phase boundary calculated directly at
  finite $\mu$. }
\end{figure}

In \Fig{fig:convrad} we show the results for both estimators for three
different truncation orders \mbox{($2n=8,16,24$)} in the ($T,\mu$)
phase diagram. Note that the difference in the radii of convergence,
estimated from $r_{2n}$ and $r^\chi_{2n-2}$ is only caused by the
prefactor in \Eq{eq:convradchi} as the same Taylor coefficients
contribute to both estimators.  Close to $T_\chi$ the oscillations in
the Taylor coefficients caused by the imaginary part of the limiting
singularity entail oscillations in the radii of convergence and do not
allow for a stable estimate of the phase boundary there.  For somewhat
smaller temperatures the radii of convergence approach the phase
boundary from above with increasing truncation order $n$. The observed
agreement in the crossover region suggests that the singularity
limiting the Taylor expansion is close to the real $\mu$-axis and the
small imaginary part is negligible. In particular, this is still valid
for the region in the vicinity of the CEP.
%%%%%%%%%%%%%%%%%%%%%%%%%%%%%%%%%%%%%%%%%%%%%%%%%%%%%%%%%%%%%%%%%%%%%%%%%%%%%
%% Pade phase boundary
%%%%%%%%%%%%%%%%%%%%%%%%%%%%%%%%%%%%%%%%%%%%%%%%%%%%%%%%%%%%%%%%%%%%%%%%%%%%%
%%%%%%%%%%%%%%%%%%%%%%%%%%%%%%%%%%%%%%%%%%%%%%%%%%%%%%%%%%%%%%%%%%%%%%%%%%%%%
%% pade convrad

The determination of the radius of convergence or the phase boundary
with the \pade{} series is more involved. For the $[N/2]$ case the
\pade{} approximant has a pole at $x=\pm \sqrt{c_{N}/c_{N+2}}$ which
coincides with the estimator $r_{2N}$, see \Eq{eq:convr}. For a
general $[L/M]$ \pade{} approximant the pole structure is more
elaborated. In the general case we use the first pole at real and
positive $\mu$ in order to estimate the phase boundary. Since all
Taylor coefficients with their corresponding error enter in the
\pade{} approximant the error propagation is more involved here in
contrast to the previous discussion where only two error sources
enter. In this context, the application of the \pade{} approximation
in lattice simulations is much more involved. However, in the model
analysis the coefficients, obtained with the AD technique, exhibit
extremely small numerical errors and result in a stable and reliable
\pade{} approximation.  In~\Fig{fig:pbpade} we show for a fixed
temperature $T=0.92\; T_\chi$ the estimates of the phase boundary
extracted from \pade{} approximations for the pressure ($p$) and the
quark number susceptibility ($\chi$) for different truncation orders
$n$. The chosen temperature is slightly above the temperature of the
CEP in the phase diagram. In addition, the estimates for the radii of
convergence ($r_{2n}$ and $r^\chi_{2n}$) and the chiral crossover
chemical potential (dashed horizontal line) are also shown. For small
truncation orders $2n \leq 8$ the $[N/2]$ \pade{} approximations
coincide with the radii of convergence as expected. For $2n >8$ we
observe that the \pade{} approximation converges faster for the
pressure as well as for the susceptibility compared to the radii of
convergence of the Taylor series.  We conclude that the \pade{}
approximation converges faster at intermediate truncation orders in
particular for the quark number susceptibility.  The estimate of the
phase boundary from the \pade{} approximation becomes comparable to
the one obtained from the Taylor expansion coefficients only at
significantly larger truncation order.  For example, in
\Fig{fig:pbpade} the distance of the \pade{} estimate to the
horizontal line at $2n=12$ is achieved with Taylor coefficients only
for $2n\geq20$. However, the lower truncation order in the \pade{}
approximation induces a more involved error propagation which might
hamper lattice simulations but not calculations with the
AD-techniques.

%%%%%%%%%%%%%%%%%%%%%%%%%%%%%%%%%%%%%%%%%%%%%%%%%%%%%%%%%%%%%%%%%%%%%%%%%%%%%
%% Locating the CEP
%%%%%%%%%%%%%%%%%%%%%%%%%%%%%%%%%%%%%%%%%%%%%%%%%%%%%%%%%%%%%%%%%%%%%%%%%%%%%

From the discussion given so far and from Fig.~\ref{fig:pbpade} one
may conclude that at fixed temperature the radius of convergence can
be estimated with an accuracy of about 15\%-20\% from a ${\cal
  O}(\mu^{12})$ series expansion. This is an acceptable uncertainty in
view of the difficulties in current QCD calculations at non-zero
chemical potential. When  $T_c$ is known the CEP could be determine
from 
\begin{equation}
\label{eq:muctc}
\mu_c = r(T_c) = \lim_{n \rightarrow \infty} r_n(T_c) \ .
\end{equation}
However, the determination of $T_c$ is still a non-trivial task. A
criterion is needed, that allows to estimate for a given expansion
order the temperature regime where all expansion coefficients may stay
positive. In this case the singularity limiting the convergence is
located on the real axis in the complex $\mu$-plane and hence
corresponds to a second-order phase transition~\cite{Stephanov:2006dnYang:1952be}.
%%%%%%%%%%%%%%%%%%%%%%%%%%%%%%%%%%%%%%%%%%%%%%%%%%%%%%%%%%%%%%%%%%%%%%%%%%%%%
%% Tc
\begin{figure}[t]
\centering
\includegraphics[width=\onefig]{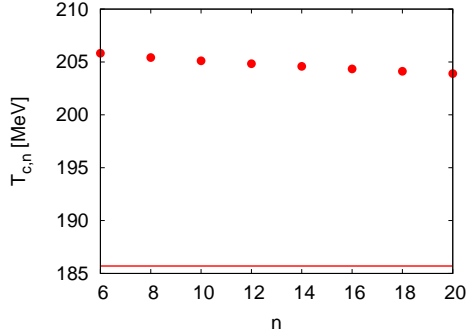}
\caption{Temperature $T_{c,n}$ at which the Taylor coefficients $c_n$
  becomes negative for different truncation orders $n$. The solid
  horizontal line denotes the critical temperature of the CEP.}
\label{fig:taylorzero}
\end{figure}
In Fig. \ref{fig:taylorzero} we plot the temperature $T_{c,n}$ where
the Taylor coefficient $c_n(T)$ becomes negative. This yields an upper
bound for the critical temperature. Of course, in the limit
$n\rightarrow \infty$ the temperature series should approach the
critical temperature of the CEP, i.e.,
\begin{equation}
T_c = \lim_{n\rightarrow \infty} T_{c,n}\ .
\end{equation}
However, we found in our model analysis only a very slow convergence
of the estimators for the critical temperature of the CEP.
Even at the 20-th truncation order there is still a temperature
difference of the order of $20$ MeV which is why we do not apply
\Eq{eq:muctc}.

%%%%%%%%%%%%%%%%%%%%%%%%%%%%%%%%%%%%%%%%%%%%%%%%%%%%%%%%%%%%%%%%%%%%%%%%%%%%%
%%
%% Summary
%%
%%%%%%%%%%%%%%%%%%%%%%%%%%%%%%%%%%%%%%%%%%%%%%%%%%%%%%%%%%%%%%%%%%%%%%%%%%%%%
\section{Summary and Conclusion}
In this talk we discussed convergence properties of the Taylor
expansion towards finite density QCD in a three flavor PQM model
framework. A newly developed differentiation technique allowed for the
calculation of high order Taylor coefficients with high precision. 
With a \pade{} resummation the number of required Taylor coefficients
can be reduced by a factor of two to obtain a similar convergence
radius.
For temperatures above the critical endpoint a good agreement of the
radius of convergence with the phase boundary is found, in particular
for the estimator of the quark number susceptibility. With a 12-th
order \pade{} approximation a satisfying estimate of the phase
boundary could be achieved. However, a determination of the critical
temperature including the location of the CEP is hampered by the slow
convergence of the estimators $T_{c,n}$.\\[2ex]

\noindent
Support by the BMBF Grant 06BI9001 and by the
Helmholtz-University Young Investigator Grant No. VH-NG-332 is
acknowledged. 
%%%%%%%%%%%%%%%%%%%%%%%%%%%%%%%%%%%%%%%%%%%%%%%%%%%%%%%%%%%%%%%%%%%%%%%%%%%%%
%% Refs
%%%%%%%%%%%%%%%%%%%%%%%%%%%%%%%%%%%%%%%%%%%%%%%%%%%%%%%%%%%%%%%%%%%%%%%%%%%%%
%\bibliography{../../../literature/qcd} 

\end{document}